# Does Using ChatGPT Result in Human Cognitive Augmentation?

Ron Fulbright[1], Miranda Morrison[2]
University of South Carolina Upstate
800 University Way, Spartanburg, SC USA 29303
[1]`fulbrigh@uscupstate.edu`
[2]`morrisme@email.uscupstate.edu`

**Abstract.** Human cognitive performance is enhanced by the use of tools. For example, a human can produce a much greater, and more accurate, volume of mathematical calculation in a unit of time using a calculator or a spreadsheet application on a computer. Such tools have taken over the burden of lower-level cognitive "grunt work" but the human still serves the role of the expert performing higher-level thinking and reasoning. Recently, however, unsupervised, deep, machine learning has produced cognitive systems able to outperform humans in several domains. When humans use these tools in a human/cog ensemble, the human's cognitive ability is augmented. In some cases, even non-experts can achieve, and even exceed, the performance of experts in a particular domain—synthetic expertise. A new cognitive system, ChatGPT, has burst onto the scene during the past year. This paper investigates human cognitive augmentation due to using ChatGPT by presenting the results of two experiments comparing responses created using ChatGPT with results created not using ChatGPT. We find using ChatGPT does not always result in cognitive augmentation and does not yet replace human judgement, discernment, and evaluation in certain types of tasks. In fact, ChatGPT was observed to result in misleading users resulting in negative cognitive augmentation.

**Keywords:** human cognitive augmentation, cognitive systems, human/cog ensembles

## 1   Introduction

Human performance of any kind is augmented by the use of tools. Physical performance is enhanced by using simple tools like hammers, shovels, and axes. Likewise, human cognitive performance is augmented by the use of tools able to process and transform information. For example, unaided, a human might take several minutes to add a column of numbers, and then the sum would need to be checked because of the possibility of error in the calculations. However, using a calculator or a computer spreadsheet, a human could produce a reliable sum in a fraction of a second. In fact, entry of the numbers becomes the limiting factor in terms of speed. Today, we commonly use software able to process words, images, video, and numbers to perform our volume of daily work. Such tools have taken over the burden of lower-level cognitive "grunt work." So far, though, the human still serves the role of the expert and performs the high-level thinking.

Recently, however, cognitive systems technology ("AIs") built using unsupervised, deep, machine learning techniques, has produced tools able to outperform humans in several domains formerly thought to be possible only as the result of high-level human cognitive processing. We call such systems "cogs." When humans use tools like these in a collaborative manner (a human/cog ensemble) human cognitive performance is enhanced—augmented. If a human's cognitive ability is augmented enough, even a non-expert can achieve, and even exceed, the performance of an expert in a particular domain, something called *synthetic expertise*. So far, though, such cognitive systems are narrow in their applicability. Even though they outperform humans, they are limited to just that domain.

Things are changing. Systems like the new Chat Generative Pre-Trained Transformer (ChatGPT), have gained much attention recently. ChatGPT is a large language model trained to predict the most probable next word in a sequence of words and is fine-tuned for conversational usage. ChatGPT mimics human-created text. Instead of being limited to a narrow domain, users can conduct extended textual dialogs with ChatGPT on practically any topic and most of the time, text generated by ChatGPT is indistinguishable from text produced by another human. Every day, millions of people use ChatGPT for assistance in learning, researching, getting advice, writing music, poetry, and prose, generating computer program code, and much more. A person using ChatGPT certainly fits our definition of a human/cog ensemble. Accordingly, the hypothesis explored in this paper is:

**H1**: In a human/cog ensemble consisting of a person using ChatGPT we should be able to observe, measure, and characterize human cognitive augmentation in the form of enhanced performance when performing a task.

To investigate the hypothesis, we designed two experiments to compare human cognitive performance with and without using ChatGPT. In one experiment we found a person using ChatGPT as a assistive tool was marginally better than a person not using ChatGPT but not enough for the result to be compelling. In the other experiment we found using ChatGPT had no effect on a person's ability to perform the task and even misled users resulting in negative cognitive augmentation.

## 2   Previous Work
### 2.1 Cognitive Systems

With recent advances in artificial intelligence (AI) and cognitive systems (cogs), we are at the beginning of a new era in human history in which humans will work in partnership with artificial entities capable of performing high-level cognition rivaling or surpassing human ability (Kelly & Hamm, 2013; Wladawsky-Berger, 2015; Gil, 2019; Fulbright, 2016a; 2016b; 2020). Already, there are artificial systems and algorithms outperforming humans and achieving expert-level results.

For example, a deep-learning algorithm has learned to detect lung cancers better than human doctors (Sandoiu, 2019). The algorithm outperforms humans in recognizing problem areas reducing false positives by 11% and false negatives by 5%.

Google's convolutional neural network, Inception v4, outperformed a group of 58 human dermatologists using dermoscopic images and corresponding diagnoses of melanoma (Haenssle et al., 2018).

In the field of diabetic retinopathy, a study evaluated the diagnostic performance of a cognitive system for the automated detection of diabetic retinopathy (DR) and Diabetic Macular Edema (DME) (Abràmoff, et al., 2018). The cog exceeded all pre-specified superiority goals.

At the University of California San Francisco and the University of California Berkeley, an algorithm running on a convolutional neural network was better than experts at finding tiny brain hemorrhages in scans of patients' heads (Kurtzman, 2019). The cog was able to complete a diagnosis in only one second, something a human would take many minutes to do.

Cognitive systems are already better than humans at diagnosing childhood depression (Lavars, 2019), predicting mortality (Wehner, 2019), detecting valvular heart disease (Stevens, 2023), and assessing cancerous tumors (Towers-Clark, 2019).

Not only are cognitive systems able to outperform humans in some domains, they are able to do things humans cannot. For example, the FIND FH machine learning model analyzed the clinical data of over 170 million people and discovered 1.3 million of them were previously undiagnosed as being likely to have familial hypercholesterolemia (Myers et al., 2019). Follow-on studies of the individual cases flagged by the cog have shown over 80% of the cases do in fact have a high enough clinical suspicion to warrant evaluation and treatment. This means on the order of 800,000 people could receive life-extending treatment who otherwise would not.

An algorithm named Word2Vec sifted through some 3.3 million abstracts and discovered associations previously unknown by human readers and predicted a new thermoelectric material four years before it was discovered (Tshitoyan, 2019; Gregory, 2019).

## 2.2 Cognitive Augmentation

We can view data, information, knowledge, and wisdom (DIKW) as a hierarchy based on relative value (Ackoff, 1989). Each level is of a higher value than the level below it partly because of the processing involved to produce the information stock at that level and partly due to the utility of the information stock at that level. Information is processed data, knowledge is processed information, etc. Processing at each level can be modeled as a cognitive process. Data, information, or knowledge, generically referred to as *information stock*, is input to the cognitive process. The cognitive process transforms the input and produces the higher-valued output. This transformation is accomplished by the expenditure of a certain amount of *cognitive work* (W) (Fulbright, 2020).

In a human/cog ensemble, a collaborative team consisting of one or more humans and one or more cognitive systems), cognitive processing of the entire ensemble is a mixture of human cognitive processing ($W_H$) and artificial cognitive processing ($W_C$) ($W^* = W_H + W_C$) as depicted in Fig. 1 (Fulbright, 2020; 2020a; Fulbright & Walters, 2020).

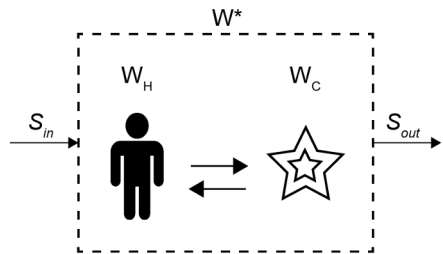

**Fig. 3.** A Human/Cog ensemble performing a cognitive process.

A human working alone is able to achieve a certain amount of cognitive work. A human aided by a cognitive system is able to achieve a greater amount of cognitive work. We call this increase in cognitive performance, *cognitive augmentation* (Fulbright, 2017; 2020).

The amount of cognitive augmentation depends on how sophisticated the cognitive system, how much of the total cognitive work it performs, and how well the human collaborates with the cognitive system. Throughout history, humans have created ever-evolving technology to assist in cognitive processing. As these systems get more capable, especially now in the era of artificial intelligence and unsupervised deep machine learning, cognitive augmentation will increase rapidly.

Different Levels of Cognitive Augmentation have been defined ranging from no augmentation at all (all human thinking) to fully artificial intelligence (no human thinking) as shown in Fig. 2 (Fulbright, 2020; 2020a; Fulbright & Walters, 2020).

| | | |
|---|---|---|
| **Level 0:** | No Augmentation | |
| | the human performs all cognitive processing | |
| **Level 1:** | Assistive Tools | |
| | e.g., abacus, calculators, software, etc. | |
| **Level 2:** | Low-Level Cognition | |
| | pattern recognition, classification, speech | |
| | human makes all high-level decisions | |
| **Level 3:** | High-Level Cognition | |
| | concept understanding, critique | |
| | conversational natural language | |
| **Level 4:** | Creative Autonomy | |
| | human-inspired/unsupervised synthesis | |
| **Level 5:** | Artificial Intelligence | |
| | no human cognitive processing | |

**Fig. 2**. Levels of Cognitive Augmentation.

In previous work, we have conducted various experiments designed to measure and characterize cognitive augmentation. Fulbright (2017; 2018) discusses several kinds of metrics and proposes several metrics to employ when measuring cognitive augmentation. Fulbright (2019) calculates cognitive augmentation for a given task finding cognitive augmentation as high as 74% when people are provided different numbers of hints by a simulated cognitive system. Fulbright & McGaha (2023) shows how information of different types affects the level of cognitive augmentation when tasked with solving several different kinds of puzzles. In both of these studies, enhanced cognitive accuracy and cognitive precision were measured.

In all three of these studies assistive information supplied to the human was simulated and did not come from an actual cognitive system. However, ChatGPT represents a cognitive system, already used by millions, with which to conduct experiments in cognitive augmentation. There have been some notable studies comparing human performance to ChatGPT.

**2.3 ChatGPT and Cognitive Augmentation**
Li et al. (2023) compared the results of ChatGPT versus human performance on the Objective Structured Clinical Examination (OSCE) in obstetrics and gynecology. ChatGPT was asked to answer discussion questions in seven key disciplines within obstetrics and gynecology. ChatGPT outscored human test-takers in questions regarding postpartum management, urogynecology and pelvic floor problems, labor management, and post-operative care. ChatGPT did not outperform humans in early pregnancy care, core surgical skills, or gynecologic oncology. Li et al. (2023) theorized those question require multiple answers and higher-level reasoning.

Kung et al. (2023) found comparable results when administering the United States Medical Licensing Examination (USMLE) to ChatGPT. ChatGPT beat the passing score of 60% on most areas but narrowly failed to pass the multiple choice-question section (59.1%) and the multiple-choice with forced justification section (52.4%) on Step 2CK of the exam. Step 2CK is typically administered to students who have successfully completed their fourth year of medical school (Kung et al., 2023).

In taking the American Board of Neurological Surgery Self-Assessment Examination 1, Ali et al. (2023) found ChatGPT 3.5 achieved a score of 73.4% and GPT-4 achieved a score of 83.4% relative to the human average of 72.8%. Both versions of ChatGPT exceeded last year's passing threshold of 69%.

Liéven et al. (2023) determined GPT-3.5 performed higher than the needed passing score on questions taken from the USMLE and MedMCQA examinations, however, GPT-3.5 still underperformed on both examinations in comparison to humans.

Brin et al. (2023), determined GPT-4 to have performed 10.77% better than human test-takers on multiple-choice questions from the USMLE involving soft skills, such as empathy, leadership, emotional intelligence, and communication.

Similarly, Eloyseph et al. (2023) found ChatGPT to score 74.35% higher than males and 66.27% higher than females on the Levels of Emotional Awareness Scale (LEAS), a test measuring emotional intelligence in the form of open-ended questions that are evaluated by licensed psychologists. It is also worth noting ChatGPT took the LEAS evaluation twice: once in January and once in February. ChatGPT improved its LEAS score by 13% between the two exams.

Duong & Solomon (2023) compared the ability of ChatGPT to humans in answering multiple-choice questions about genetics. Humans answered questions with 1.61% greater accuracy overall, but ChatGPT performed 8.51% better than humans on questions that relied on memorization instead of critical thinking.

Jarou et al. (2023) administered multiple-choice questions from the American College of Emergency Physicians (ACEP) study guide. Human respondents scored 36.32% higher than GPT-3.5 and 19.5% higher than GPT-4; yet GPT-4 scored 33.71% higher than GPT-3.5.

Additionally, Katz et al. (2023) compared the performance of GPT-3.5 and GPT-4 with human respondents on the Uniform Bar Exam. GPT-4 performed 11.3% better than human test-takers, but GPT-3.5 could not exceed human performance in any of the subject areas tested on the exam.

Instead of situations in which ChatGPT replaces humans, this paper is interested in exploring how using ChatGPT enhances a user's cognitive ability as stated in the hypothesis, H1. Unfortunately, there have not yet been a lot of studies like this. Noy and Zhang (2023) assigned 444 professionals with tasks related to their respective profession, such as sensitive e-mails, press releases, and reports. After completing the initial task, the group was split in two. The control group was asked to repeat the task using LaTeX, a document preparation program, while the test group used ChatGPT to assist them with their second task. Those using ChatGPT reduced the time spent on the task by 35.16% and improved their score on the second task by 15.45%.

## 3    The Experiments

This paper presents the results of two experiments in which we asked students to perform a task and comparing their performance with that of an expert. Students were tasked with two different challenges, an innovation problem and an expert advice question. Students used ChatGPT 3.5 (circa November 2023, January 2024) in these experiments.

### 3.1 Innovation Challenge

For the innovation challenge, students were given the following problem statement:

> *When shooting skeet, fragments from the skeet fall on and cause harm to the grass field by preventing sunlight and water from reaching the grass. What changes can I make to protect the grass?*

This is the same problem statement used in a previous cognitive augmentation experiment described in Fulbright (2019). In that experiment, students were given hints in the form of innovative suggestions (called *operators*) and the results showed it was possible to affect the innovative solutions arrived at by the participants toward a desired goal—the preferred solution. In fact, results showed as much of a 74% increase in cognitive accuracy was achieved demonstrating significant cognitive augmentation.

Participants were asked to synthesize three innovative solutions to the problem. Any solution could not interfere with the sheet shooting activity, must be relatively easy and inexpensive to implement, and involve as little change to the current situation as possible. One-half of the participants were instructed to not use any Internet-based resource at all. The other half were instructed to use only ChatGPT.

This innovation problem was chosen because it is a problem used in teaching innovation at the university undergraduate level for over many years. As such, there is a long history of solutions, and patterns of solutions, to compare new results with. Because of this history, we know what type of solutions people give when not aided by any cognitive system or assistive information and we know what type of solutions are given by professional/expert innovators.

With respect to H1, our goal was to see if using ChatGPT altered the type of solutions. If H1 was verified in this experiment, we would expect to see the solutions trend toward the professional/expert type of solutions.

**3.2 Retirement Decision**

For the second experiment, participants were given detailed information about an imaginary college professor approaching retirement. Information provided included: age, profession (and pros and cons of the profession), salary, debt, medical situation, retirement savings, with the goals of being able to remain in the current home, travel at least twice per year after retirement, and not outliving their money.

Participants were asked if the person should retire early at 67 or wait until the age of 70. A person can go to a retirement planning expert and ask this question and receive a detailed response including an explanation of why it is better to retire at 67 or wait until 70. None of the participants, being university students, were experts in retirement planning. However, we asked each participant to provide a specific answer (either 67 or 70) and then also provide a justification to support the answer. In our judgement of the results, it did not matter which age was given as the answer. We focused on the level of detail in the justification. A detailed and specific justification, in our view, constituted an expert-level answer to the challenge.

One-half of the participants were instructed to use ChatGPT only, an no other Internet-based resource, and the other half was instructed to use any Internet-based resource except ChatGPT. With respect to H1, we expected to see an increase in the ability of participants to provide an expert-level answer due to using ChatGPT.

## 4 The Results
### 4.1 Innovation Challenge Results

For the innovation challenge, 13 students used ChatGPT and 13 students did not use ChatGPT to synthesize a total of N=96 ideas to solve the skeet shooting innovation challenge. As we have seen in earlier studies using this problem statement, ideas fell into three broad categories: changing the field (**F**), changing the skeet (**T**), or ideas not solving the problem at all (**X**). Ideas involving the field fell into three different subcategories and ideas involving the skeet fell into two categories:

- $F^T$     protecting the field with a tarp, net, or some other kind of covering
- $F^C$     ways of cleaning the field or making picking up fragments easier
- $F^G$     changing or replacing the grass on the field
- $T^B$     replacing clay skeet with biodegradable material
- $T^C$     changing the clay skeet to make cleanup more easier
- **X**     ideas addressing ideas other than the stated problem

Fig. 4 shows the results for students not using ChatGPT. Overwhelmingly, most ideas (79.5%) involved changing the field in some way such as covering it with a tarp or net to prevent fragments from reaching the grass or various ways to clean the field after fragments have fallen onto the grass. The remainder (20.5%) of the ideas involved changing the clay skeet such as making the skeet out of biodegradable material or out of some material other than clay to facilitate easier cleanup. Field-related ideas out numbered skeet-related ideas 3.8:1.

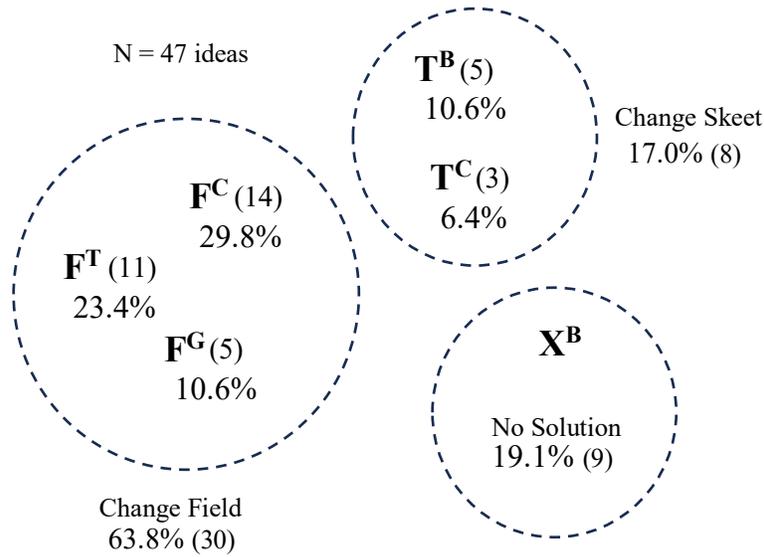

**Fig. 3**. Solutions Using ChatGPT for Assistance (47 ideas).

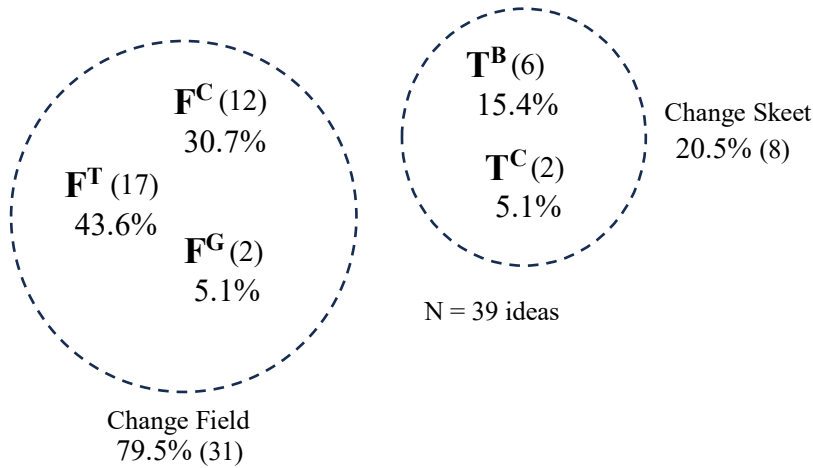

**Fig. 4**. Solutions Not Using ChatGPT for Assistance (39 ideas).

Fig. 3 shows the results for students using ChatGPT. As in Fig 4, ideas involving changing the field vastly outnumber ideas involving changing the skeet by almost exactly the same ratio 3.75:1. Therefore, we see no difference in the type of ideas generated as a result of using ChatGPT. Therefore, the hypothesis, H1, is refuted.

Interestingly, students using ChatGPT synthesized a number of ideas having no effect at all on the primary problem—littering of the grass by the fragments. Examining these ideas in detail shows these ideas were related to "educating shooters about the environmental impact" and "educating shooters about gun safety." These ideas can be explained when one analyzes the response from ChatGPT when given the problem statement as the prompt. ChatGPT is trained from articles and other content available on the Internet. Because the problem statement involves guns and shooting, ChatGPT responded with suggestions to educate shooters about gun safety because on the Internet, when one sees a document about guns and shooting, it is very likely to also include comments about safety. Even though the concepts of guns and safety are understandably related, the safety issue has nothing to do with solving the problem given in the problem statement—littering the grass field. ChatGPT however does not perform such in-depth analysis to realize this. ChatGPT's responses are driven by word association. Likewise, because the problem statement mentions littering and damaging grass, ChatGPT finds associations with environmental issues important and therefore responded to students suggesting education about the environment since this is found in millions of pages on the Internet when litter and harming grass is mentioned. While one could argue you might be able to talk a shooter out of shooting after they understand the harm to the grass, this is not likely to change the mind of the vast majority of shooters, so is not a practical solution. Interestingly, in this case, using of ChatGPT actually distracted students by misleading them to consider things having nothing to do with the problem. Therefore, one could argue using ChatGPT actually decreased cognitive ability—resulting in *negative cognitive augmentation.*

### 4.2 Retirement Decision Results

For the retirement decision challenge, 15 students used ChatGPT and 10 students did not use ChatGPT. The challenge asked students to provide a specific answer, whether or not the subject should retire at 67 or 70 and also provide an expert-level justification of that answer. We explained to the students how people could visit a retirement planning professional and receive guidance and we asked students to provide a similar-quality answer here.

Responses were judged to be either "expert quality" or "non-expert quality" as seen in Fig. 5. The difference between an expert and a non-expert response is in the details provided in the justification. To answer the question properly, one must calculate the monthly inflow and outflow of money. To do that, one has to find out how much per month social security payments would be and add to that withdrawals from savings to augment the monthly inflow. Once this is established, one has to calculate how long the subject's money would last. Very different answers are obtained if one retires at age 67 versus 70. In judging the responses, we did not consider which answer the student provided. It did not matter at what age the student decided the subject should retire. What we did look for, though, is did the student conduct and include the analysis needed to justify their response. Reponses including the analysis were deemed "expert" and the responses not including the analysis were deemed "non-expert."

Another characteristic of non-expert responses was "generic" information like "the person must consider how long their savings will last." While this is certainly is something a person needs to consider when planning retirement, one would not have to visit an expert to get this advice. Any friend, family member, or easy search on the Internet will produce a list of such things for one to consider. In fact, the first response from ChatGPT gives a list of 8-10 such generic issues to consider. So, a response simply containing generic information like this was considered "non-expert."

Fig. 5 shows students not using ChatGPT provided expert-quality answers 40% of the time. Students using ChatGPT provided expert-quality answers 53% of the time. While this is an increase, it is not a definitive increase in our opinion. Of further note, is of the students using ChatGPT, there was only one more expert response than non-expert response. If ChatGPT provided demonstrable cognitive augmentation for this task, one would expect many more expert answers than non-expert answers from the group of students using ChatGPT.

Students not using ChatGPT were allowed to use any other Internet-based resource and reported the tool or information source they used. We observed all expert-quality answers from the non-ChatGPT group were provided by students who used a retirement calculator available on the Internet. We believe the students using a retirement calculator were cognitively augmented just like students using ChatGPT. In fact, the retirement calculator is an assistive tool designed specifically to help answer retirement planning questions whereas ChatGPT is not. Although we are not able to definitively conclude it in this study, we believe if non-ChatGPT students were not allowed to use a retirement calculator, the number of expert-quality answers would be much lower and students using ChatGPT would have performed much better.

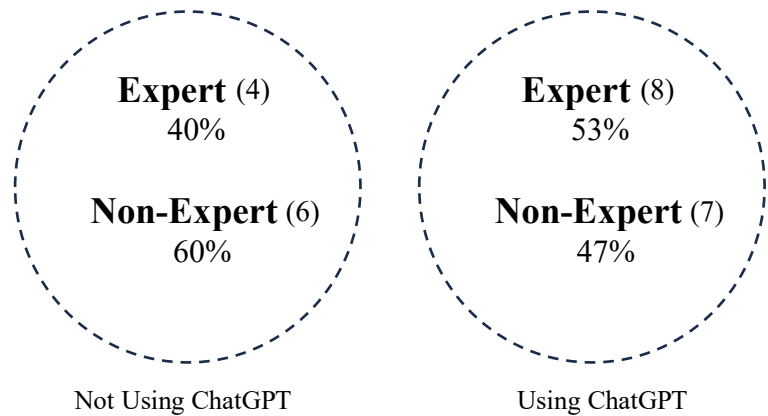

**Fig. 5**. Responses to the Retirement Decision Challenge (25 responses).

## 5   Conclusion

Our hypotheses, H1 was refuted in the innovation experiment and only moderately confirmed in the retirement decision experiment. In fact, in the innovation experiment, ChatGPT actually misled students to thinking about issues irrelevant to the problem statement, resulting in negative cognitive augmentation. Both experiments involved tasks requiring detailed analysis, high-level reasoning, and human judgment and were questions without a definite right and wrong answer. To this extent, we confirm the findings of Li et al. (2023), Liéven et al. (2023), Jarou et al. (2023), and Katz et al. (2023) who found ChatGPT outperformed humans on some types of questions but not those involving higher-level analysis.

Our results show using ChatGPT does not guarantee expert-level performance. None of the students participating in this study were experts at using ChatGPT. For some, this task was the first time they ever used ChatGPT. If students had more experience with ChatGPT, more expert-level results might be expected. Also, students who participated were not given detailed instructions on how to answer retirement questions nor how to think innovatively. If they had knew more about the subject, it stands to reason more would have been able to provide expert-quality answers. This can be explored in future studies.

It is necessary to note, when designing these experiments, we found it quite difficult to determine tasks to give to students. We tested and discarded several tasks before deciding on the innovation and retirement challenges because we found ChatGPT was able to simply spit out a perfectly correct answer on the first prompt. Over time, we realized we could not ask students to perform any task involving just simple knowledge retrieval because ChatGPT does this quite well. To create a challenge tough enough, we realized the tasks needed to require cognitive processes involving *understanding*, *evaluation*, *appraisal, critique*, and *judgment* in order to exercise the students and ChatGPT more vigorously.

We recognize these types of cognitive processes represent the upper levels of Bloom's Taxonomy, a framework for categorizing educational goals and therefore classifying levels of cognitive processes (Bloom et. al., 1956; Anderson & Kratwohl, 2001). We expect future studies to show ChatGPT already able to take the cognitive "grunt work" of lower-level cognitive processes like *recall, defining, listing, classifying, describing, discussing, explaining, translating,* and *recognizing* away from the human in a human/cog ensemble. Any task involving these levels of cognitive processing will be done quicker and better by ChatGPT leaving the human to do the higher-level cognitive processing. Relieving the human of the cognitive "grunt work" will result in significant cognitive augmentation in the form of higher-quality, higher-value results in less time with less effort.

## References

1. Abràmoff, M. D., Lavin, P. T., Birch, M., Shah, N. and Folk, J. C. (2018). Pivotal trial of an autonomous AI-based diagnostic system for detection of diabetic retinopathy in primary care offices. *Digital Med*., 1: 39. Available online at: https://www.nature.com/articles/s41746-018-0040-6, last accessed January 2024.

2. Ackoff, R. (1989). From Data to Wisdom, *Journal of Applied Systems Analysis,* 16. Available online: https://faculty.ung.edu/kmelton/Documents/DataWisdom.pdf last accessed February 2023.

3. Ali, R., Tang, O. Y., Connolly, I. D., Zadnik Sullivan, P. L., Shin, J. H., Fridley, J. S., Asaad, W. F., Cielo, D., Oyelese, A. A., Doberstein, C. E., Gokaslan, Z. L., & Telfeian, A. E. (2023). Performance of CHATGPT and GPT-4 on Neurosurgery Written Board Examinations. Neurosurgery. https://doi.org/10.1227/neu.0000000000002632.

4. Anderson, L. W., Krathwohl, D. R., Airasian, P. W., Cruik-shank, K. A., Mayer, R. E., (2001). A taxonomy for learning, teaching, and assessing: A revision of Bloom's taxonomy of educational objectives, Pearson.

5. Bloom, B. S.,  Engelhart, M. D., Furst, E. J., Hill, W. H., Krath-wohl, D. R., (1956). Taxonomy of educational objectives: The classification of educational goals. Handbook I: Cognitive domain. New York: David McKay Company.

6. Brin, D., Sorin, V., Vaid, A., Soroush, A., Glicksberg, B. S., Charney, A. W., Nadkarni, G., & Klang, E. (2023). Comparing ChatGPT and GPT-4 Performance in USMLE Soft Skill Assessments. Scientific Reports, 13(1). https://doi.org/10.1038/s41598-023-43436-9.

7. Elyoseph, Z., Hadar-Shoval, D., Asraf, K., & Lvovsky, M. (2023). ChatGPT Outperforms Humans in Emotional Awareness Evaluations. Frontiers in Psychology, 14. https://doi.org/10.3389/fpsyg.2023.1199058.

8. Fulbright, R. (2016a). The Cogs Are Coming: The Cognitive Augmentation Revolution, *Proceedings of the Association Supporting Computer Users in Education 2015* (49[th], Myrtle Beach, SC). Available online at: https://eric.ed.gov/?q=cognitive+development+ in+early+childhood&ff1=dtysince_2014&pg=1340&id=ED570900, last accessed January 2024.

9. Fulbright, R. (2016b). How Personal Cognitive Augmentation Will Lead To The Democratization Of Expertise, *Advances in Cognitive Systems*, 4. Available online at: http://www.cogsys.org/posters/2016/poster-2016-3.pdf, last viewed January 2024.

10. Fulbright, R. (2017). Cognitive Augmentation Metrics Using Representational Information Theory, In: Schmorrow D., Fidopiastis C. (eds) *Augmented Cognition. Enhancing Cognition and Behavior in Complex Human Environments,*  AC 2017, *Lecture Notes in Computer Science*, vol 10285. Springer, Cham. Available online: https://link.springer.com/chapter/10.1007/978-3-319-58625-0_3 last accessed 2023.

11. Fulbright, R. (2018). On Measuring Cognition and Cognitive Augmentation, In: Yamamoto, S., Mori, H. (eds) *Human Interface and the Management of Information*, HIMI 2018, *Lecture Notes in Computer Science*, vol 10905. Springer, Cham. Available online: https://link.springer.com/chapter/10.1007/978-3-319-92046-7_41 last accessed February 2023.

12. Fulbright, R. (2019). Calculating Cognitive Augmentation -A Case Study, In: Schmorrow, D. and Fidopiastis, C. (eds) *Augmented Cognition, AC 2019, Lecture Notes in Computer Science*, vol 11580. Springer, Cham. Available online: https://link.springer.com/chapter/10.1007/978-3-030-22419-6_38 last accessed February 2023.

13. Fulbright, R. (2020). *Democratization of Expertise: How Cognitive Systems Will Revolutionize Your Life,* CRC Press, Boca Raton, Fl.

14. Fulbright, R. (2020a). The Expertise Level, In: Schmorrow D., Fidopiastis C. (eds) *Augmented Cognition. Human Cognition and Behavior*. HCII 2020. *Lecture Notes in Computer Science*, vol 12197. Springer, Cham. Available online: https://link.springer.com/chapter/10.1007/978-3-030-50439-7_4 last accessed February 2023.

15. Fulbright R. and Walters, G. (2020). Synthetic Expertise, In: Schmorrow D., Fidopiastis C. (eds) *Augmented Cognition. Human Cognition and Behavior*. HCII 2020. *Lecture Notes in Computer Science*, vol 12197. Springer, Cham. Available online: https://link.springer.com/chapter/10.1007/978-3-030-50439-7_3 last accessed February 2023.

16. Fulbright R. and McGaha, S. (2023). The Effect of Information Type on Human Cognitive Augmentation, In: Schmorrow, D. and Fidopiastis, C. (eds), *Augmented Cognition: 17th International Conference, AC 2023*, held as Part of the 25th HCI International Conference, HCII 2023, Copenhagen, Denmark, July 23–28, 2023, pages 206- 220. Available online at: https://dl.acm.org/doi/abs/10.1007/978-3-031-35017-7_14, last viewed January 2024.

17. Gil, D. (2019). Cognitive systems and the future of expertise, YouTube video located at https://www.youtube.com/watch?v=0heqP8d6vtQ and last accessed February 2023.

18. Gregory, M., (2019). AI Trained on Old Scientific Papers Makes Discoveries Humans Missed, *Vice* Internet page located at: https://www.vice.com/en_in/article/neagpb/ai-trained-on-old- last accessed January 2024.

19. Haenssle, H. A., Fink, C., Schneiderbauer, R., Toberer, F., Buhl, T., Blum, A., Kalloo, A., Hassen, A. B. H., Thomas, L., Enk, A. and Uhlmann, L. (2018). Man against machine: diagnostic performance of a deep learning convolutional neural network for dermoscopic melanoma recognition in comparison to 58 dermatologists. *Annals of Oncology*, 29(8): 1836–1842. August. Available online: https://academic.oup.com/annonc/article/29/8/1836/5004443, last accessed November 2019.

20. Jarou, Z. J., Dakka, A., McGuire, D., & Bunting, L. (2023). ChatGPT Versus Human Performance on Emergency Medicine Board Preparation Questions. Annals of Emergency Medicine. https://doi.org/10.1016/j.annemergmed.2023.08.010.

21. Katz, D. M., Bommarito, M. J., Gao, S., & Arredondo, P. (2023). GPT-4 Passes the Bar Exam. SSRN Electronic Journal. https://doi.org/10.2139/ssrn.4389233.

22. Kelly, J.E. and Hamm, S. (2013). *Smart Machines: IBMs Watson and the Era of Cognitive Computing*, Columbia Business School Publishing, Columbia University Press, New York, NY.

23. Kung, T. H., Cheatham, M., Medenilla, A., Sillos, C., De Leon, L., Elepaño, C., Madriaga, M., Aggabao, R., Diaz-Candido, G., Maningo, J., & Tseng, V. (2023). Performance of ChatGPT on USMLE: Potential for AI-Assisted Medical Education Using Large Language Models. PLOS Digital Health, 2(2). https://doi.org/10.1371/journal.pdig.0000198

24. Kurtzman, L. (2019). AI Rivals Expert Radiologists at Detecting Brain Hemorrhages: Richly Annotated Training Data Vastly Improves Deep Learning Algorithm's Accuracy, *UCSF News*, University of California San Francisco. Available online at: https://www.ucsf.edu/news/2019/10/415681/ai-rivals-expert-radiologists-detecting-brain-hemorrhages, last viewed January 2024.

25. Lavars, N. (2019). Machine learning algorithm detects signals of child depression through speech, *New Atlas,* published May 7. Available online at https://newatlas.com/machine-learning-algorithm-depression/59573/ last accessed February 2023.

26. Li, S. W., Kemp, M. W., Logan, S. J. S., Dimri, P. S., Singh, N., Mattar, C. N. Z., Dashraath, P., Ramlal, H., Mahyuddin, A. P., Kanayan, S., Carter, S. W. D., Thain, S. P. T., Fee, E. L., Illanes, S. E., Choolani, M. A., Rauff, M., Biswas, A., Low, J. J. H., Ng, J. S., … Lim, M. Y. (2023). ChatGPT Outscored Human Candidates in a Virtual Objective Structured Clinical Examination in Obstetrics and Gynecology. *American Journal of Obstetrics and Gynecology*, 229(2). Available online at: https://pubmed.ncbi.nlm.nih.gov/37088277/, last viewed January 2024.

27. Liévin, V., Hother, C. E., & Winther, O. (2023). Can Large Language Models Reason About Medical Questions? arXiv. https://doi.org/https://doi.org/10.48550/arXiv.2207.08143.

28. Myers, K. D., Knowles, J. W., Staszak, D., Shapiro, M. D., Howard, W., Yadava, M., Zuzick, D., Williamson, L., Shah, N. H., Banda, J. M., Leader, J., Cromwell, W. C., Trautman, E., Murray, M. F., Baum, S. J., Myers, S., Gidding, S. S., Wilemon, K. and Rader, D. J. (2019). Precision screening for familial hypercholesterolaemia: a machine learning study applied to electronic health encounter data, *Lancet Digital Health*, Available online at: https://www.thelancet.com/journals/ landig/article/PIIS2589-7500(19)30150-5/fulltext, last accessed January 2024.

29. Noy, S., and Zhang, W. (2023). Experimental Evidence on the Productivity Effects of Generative Artificial Intelligence, *Science*, Vol 381, Issue 6654, pp. 187-192. Available online at: https://www.science.org/doi/10.1126/science.adh2586, last viewed January 2024.

30. [Stevens] (2023). Listen to your heart: AI tool detect heart diseases that doctors often miss, *Stevens Institute of Technology* media release. Available online at: https://www.stevens.edu/ news/listen-to-your-heart-ai-tool-detects-cardiac-diseases-that-doctors-often#, last viewed January 2024.

31. Sandoiu, A. (2019). Artificial intelligence better than humans at spotting lung cancer. *Medical News Today Newsletter*, May 20. Available online: https://www.medicalnewstoday.com/articles/325223.php#1, last accessed November 2019.

32. Tshitoyan, V., Dagdelen, J., Weston, L., Dunn, A., Rong, Z., Kononova, K., Persson, A., Ceder, G., and Jain, A. (2019). Unsupervised word embeddings capture latent knowledge from materials science literature, *Nature*, 571. Available online at: https://www.nature.com/articles/s41586-019-1335-8, last viewed January 2024.

33. Towers-Clark, C. (2019). The Cutting-Edge of AI Cancer Detection, *Forbes,* published April 30. Available online at https://www.forbes.com/sites/charlestowersclark/2019/04/30/the-cutting-edge-of-ai-cancer-detection/#45235ee77336 last accessed February 2023.

34. Wehner, M. (2019). AI is now better at predicting mortality than human doctors, *New York Post,* published May 14. Available online at https://nypost.com/2019/05/14/ai-is now-better-at-predicting-mortality-than-human doctors/?utm_campaign=partnerfeed&utm_medium=syndicated&utm_source=flipboard last accessed February, 2023.

35. Wladawsky-Berger, I. (2015). The Era of Augmented Cognition, *The Wall Street Journal: CIO Report* Internet page located at http://blogs.wsj.com/cio/2013/06/28/the-era-of-augmented-cognition/ last accessed February 2023.